\documentclass[aps,pra,twocolumn,showpacs,preprintnumbers,amsmath,letterpaper]{revtex4}

\usepackage{graphicx}
\usepackage{dcolumn}
\usepackage{bm}

\newcommand{\bp}[1]{\left( #1 \right)}
\newcommand{\bs}[1]{\boldsymbol{#1}}

\newcommand{\ket}[1]{\ensuremath{| #1 \rangle}}
\newcommand{\av}[1]{\left< #1 \right>}
\newcommand{\nn}{\nonumber}

\begin{document}

\title{Influence of Pure Dephasing\\on Emission Spectra from Single Photon Sources}

\author{A. Naesby}
\author{T. Suhr}
\author{P. T. Kristensen}
\author{J. M\o rk}
\email{jesm@fotonik.dtu.dk}

\affiliation{Department of Photonics Engineering, Technical University of Denmark (DTU), Building 345 V, DK-2800 Kgs. Lyngby, Denmark}

\date{\today}

\begin{abstract}
We investigate the light-matter interaction of a quantum dot with the electromagnetic field in a lossy microcavity and calculate emission spectra for non-zero detuning and dephasing. It is found that dephasing shifts the intensity of the emission peaks for non-zero detuning. We investigate the characteristics of this intensity shifting effect and offer it as an explanation for the non-vanishing emission peaks at the cavity frequency found in recent experimental work.
\end{abstract}
\maketitle

\begin{figure}
	[!b]
	\begin{center}	
	\includegraphics[width=8.6cm]{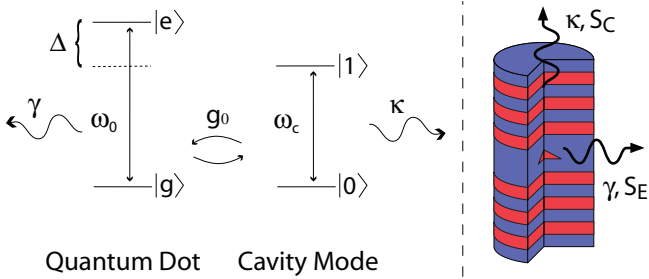} 
	\caption{\label{fig:ModelSystem}(Color online) Left: Schematic displaying the energy levels of the two-level QD and cavity. \ket{e} and \ket{g} denote excited and ground state of the emitter and \ket{1} and \ket{0} denote the excited and empty cavity mode. Right: Schematic of a micropillar with a QD in a high-Q cavity. Light escapes from the cavity in the forward direction at a rate $\kappa$, while the QD excitation decays at rate $\gamma$.}
	\end{center}
\end{figure}
The realization of a solid-state single photon source has been given much attention, because of the many potential applications for such a device. The particularly promising scheme, where a Quantum Dot (QD) is coupled to a high-Q microcavity \cite{Yamamoto93, Vahala03}, has been investigated both experimentally \cite{Reithmaier04, Yoshie04, Hennessy07} and theoretically \cite{Carmichael89, Cui06}. Recent experimental results show a significant emission at the cavity resonance even for strongly detuned systems \cite{Reithmaier04, Yoshie04, Hennessy07}, which is not well understood. In order to understand the physics and limitations, it is of significant interest to develop detailed models for such structures, that rely on the coupling between a two-level emitter and a cavity mode resonance. The role of dephasing in QD systems was pointed out by Cui and Raymer \cite{Cui06}, who showed that pure dephasing broadens the emission peaks and softens the features of the emission spectra from a resonantly coupled QD-cavity system. We extend the results of Cui and Raymer to the realistic case of non-zero detuning between cavity and QD resonance and show that detuned systems display a surprisingly large dephasing dependence, which leads to an intensity shift similar to recent experimental observations \cite{Reithmaier04, Yoshie04, Hennessy07}.

We consider the model of Cui and Raymer \cite{Cui06}, indicated in fig. \ref{fig:ModelSystem}, where a QD emitter and a cavity are treated as coupled two-level systems with coupling strength $g_0$. Both QD and cavity couple to output reservoirs, so that photons escape from the cavity at a rate $\kappa$ and the and the excitation of the QD decays nonradiatively and to other modes a a total rate $\gamma$. The resonance frequencies of the QD and cavity mode are denoted $\omega_0$ and $\omega_c$ respectively, and $\Delta = \omega_0-\omega_c$ is the detuning.

The interaction hamiltonian is found for the quantized field in the rotating wave approximation and is given in the interaction picture by \cite{Cui06, Scully97}
\begin{eqnarray}
H_I &=& \hbar g_0 \sigma_+ae^{i\Delta t} + \hbar\sum_{\bs{p}} A_{\bs{p}}^*\sigma_-d_{\bs{p}}^\dagger e^{i\delta_pt} \nn \\
&+& \hbar\sum_{\bs{k}} B_{\bs{k}}^*ab_{\bs{k}}^\dagger e^{i\delta_kt} + \mathrm{H.c.} \label{eq:interaction_hamiltonian}
\end{eqnarray}
where $\sigma_\pm$ are the raising/lowering operators of the QD and  $a^{\bp{\dagger}}$, $b^{\bp{\dagger}}$, $d^{\bp{\dagger}}$ are cavity, cavity reservoir and QD reservoir lowering (raising) operators obeying bosonic statistics. $A_{\bs{p}}^{\bp{*}}$ and $B_{\bs{k}}^{\bp{*}}$ are coupling strengths  for the interaction with the $p$'th QD reservoir mode and the $k$'th cavity reservoir mode and $\delta_p = \omega_p-\omega_0$ and $\delta_k = \omega_k-\omega_c$ are detunings for the QD output reservoir and cavity output reservoir, respectively. The last two terms in eqn. (\ref{eq:interaction_hamiltonian}) describe the coupling to emitter and cavity output, respectively.

The system is initiated with an excitation of the emitter and is described by the state vector
\begin{equation}
\ket{\Psi} = E\ket{e,0} + C\ket{g,1} + \sum_{\mathbf{p}} E^r_\mathbf{p} \ket{g,\mathbf{p}} + \sum_{\mathbf{k}}C^r_\mathbf{k} \ket{g,\mathbf{k}} \label{eq:State}
\end{equation}
where $|E(t)|^2$ and $|E^r_{\mathbf{p}}(t)|^2$ ($|C(t)|^2$ and $|C^r_{\mathbf{p}}(t)|^2$) are slowly varying probability amplitudes for the emitter and emitter decay reservoir (cavity and cavity decay reservoir), respectively. By inserting eqns. (\ref{eq:interaction_hamiltonian}) and (\ref{eq:State}) into the Schr\"odinger equation, the envelope functions are extracted by projecting onto the different states of the system and are given by
\begin{eqnarray}
\partial_t E\bp{t} &=& -ig_0e^{+i\Delta t}C\bp{t} - \gamma E\bp{t} \label{IV_diff_E1}\\
\partial_t C\bp{t} &=& -ig_0e^{-i\Delta t}E\bp{t} - \kappa C\bp{t} \label{IV_diff_C1}
\end{eqnarray}
where the Wigner-Weisskopf approximation has been employed to transform the reservoir coupling terms in the Hamiltonian into the decay terms $\kappa$ and $\gamma$.
Dephasing is modeled as a random gaussian process as in \cite{Cui06, Mandel97} and included in eqns. (\ref{IV_diff_E1}) and (\ref{IV_diff_C1}) by letting $\omega_0 t \rightarrow \omega_0 t + \int_0^t dt f\bp{t}$, where $f\bp{t}$ is a stochastic Langevin noise force with characteristics $\left< f\bp{t} \right> = 0$ and $\left< f\bp{t}f\bp{t'} \right> = 2\gamma_p \delta\bp{t-t'}$, where $\gamma_p$ is the dephasing rate \cite{Haken04}. By introducing dephasing only in the coupling to the cavity mode, \cite{Cui06}, we neglect dephasing induced broadening of the, assumed weak, emission to other (leaky) modes.

Following \cite{Cui06, Wod79}, eqns. (\ref{IV_diff_E1}) and (\ref{IV_diff_C1}) (with dephasing included) are transformed into simpler equations of motion for $E(t)$ and $C(t)$ and finally solved in order to extract the emission spectra, given by
\begin{eqnarray}
S_{E} &=& \frac{2\gamma}{\pi} \Re \left\{\int_0^\infty e^{i\bp{\Omega-\Delta} \tau} \av{E\bp{t+\tau}E^*\bp{t}} d\tau dt \right\} \label{eq:emitter_emission} \\
S_{C} &=& \frac{2\kappa}{\pi} \Re \left\{\int_0^\infty e^{i\Omega \tau} \av{C\bp{t+\tau}C^*\bp{t}} d\tau dt \right\} \label{eq:cavity_emission}
\end{eqnarray}
where $S_E$ and $S_C$ are the emission spectra for the emitter and cavity, respectively, and $\Omega$ is the frequency of the emitted light. The emission spectra characterize the light that escapes the QD-cavity system through the decay rates $\gamma$ and $\kappa$ and corresponds to what is measured in photoluminescence experiments. In general the measured spectrum is expected to be a combination of $S_E$ and $S_C$ depending on the geometry and exact details of the setup. This is because the emitter can couple to both the cavity mode and to radiation modes outside the cavity and in a photoluminescence experiment, the detector picks up emission from both the cavity and the emitter. For highly directional micropillar type setups, as the one shown in fig. \ref{fig:ModelSystem}, the cavity emission is expected to dominate the measured light, whereas the emitter spectrum becomes important in photonic crystal QD-cavities, as has been suggested by Auffeves et al. \cite{auffeves-2007}.
\begin{figure}
	[!tbhp]
	\begin{center}
	   \includegraphics[width=8.6cm]{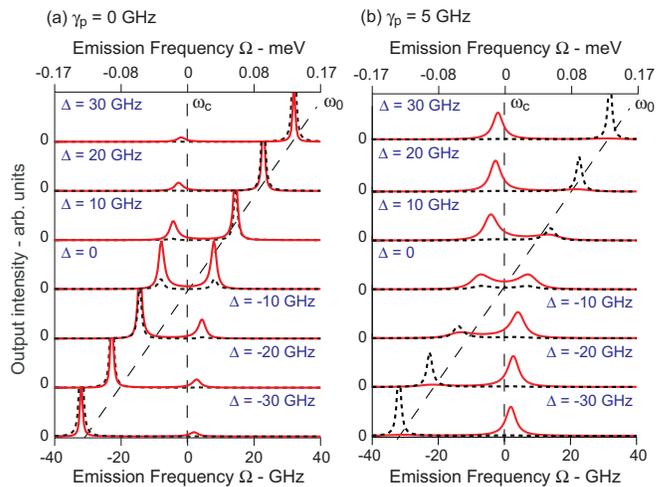} 
        \caption{\label{fig:Spectrum-Comb}(Color online) $S_C$ (full) and $S_E$ (dashed) for (a)~zero~dephasing and (b)~5~GHz dephasing rate. The emitter ($\omega_0$) and cavity ($\omega_c$) frequencies are indicated with the dashed lines and the total emission intensity $\int d\Omega \bp{S_{C} + S_{E}}$ is constant for all detunings. \mbox{\textit{Parameters: }$g_0=8$~GHz, $\kappa=1.6$~GHz and $\gamma=0.32$~GHz.}}
	\end{center}
\end{figure}
In fig. \ref{fig:Spectrum-Comb} we show $S_C$ (full line) and $S_E$ (dashed line) for a dephasing rate of zero (a) and 5~GHz (b). The parameters are chosen so that the system is in the strong coupling regime with $g_0 = 8$~GHz, $\kappa = 1.6$~GHz and $\gamma =0.32$~GHz, and the Rabi oscillations lead to a splitting of the emission peaks when the emitter and cavity are resonant \cite{Michler04, Inoue08}. The anti-crossing characteristic of strong-coupling is clearly seen in fig. \ref{fig:Spectrum-Comb}. In this context we define the strong coupling regime as $g_0 > \kappa, \gamma$. The general definition \cite{Michler04} also contains the detuning $\Delta$ and far detuned systems are thus not necessarily in the strong-coupling regime. For zero dephasing (fig. \ref{fig:Spectrum-Comb} (a)) it is noted how the peak at the emitter frequency dominates both $S_E$ and $S_C$ at high detuning, which is a result of the decrease of the coupling as the detuning is increased and of starting the system with an excitation of the emitter.

The inclusion of dephasing (fig. \ref{fig:Spectrum-Comb} (b)) considerably changes the emission spectra: First, the peaks are broadened and the splitting originating from Rabi oscillations is blurred, as has already been shown in \cite{Cui06}. Secondly, and more surprising, the inclusion of dephasing for non-zero detuning leads to a qualitative change in the cavity spectrum $S_C$ as dephasing shifts the emission intensity toward the cavity frequency.

\begin{figure}
	[!tb]
	\begin{center}
		\includegraphics[width=8.6cm]{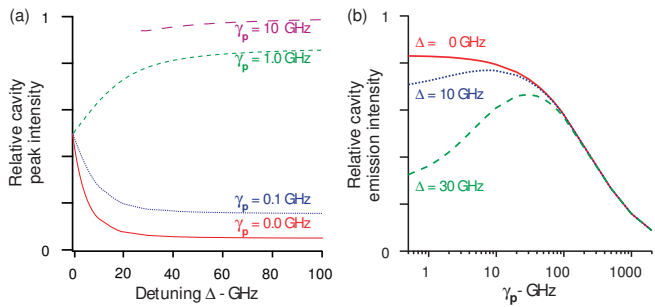} 
        \caption{\label{fig:Emission}(Color online) (a) Relative left peak (cavity peak) intensity in $S_C$ as a function of detuning $\Delta$. (b) Relative cavity emission intensity as a function of dephasing rate $\gamma_p$. \textit{Parameters:} $g_0=8$~GHz, $\kappa=1.6$~GHz, $\gamma=0.32$~GHz. }
	\end{center}
\end{figure}

This intensity shifting effect is present both in the strong and weak coupling regime as well as for very large detunings $(|\Delta| \gg g_0)$, but only in the cavity emission spectrum. The intensity shifting effect is illustrated in fig. \ref{fig:Emission}  (a) where the peak intensity (i.e. the maximum output value in a narrow interval around the peak) of the leftmost peak in $S_C$ (which becomes identical to the cavity emission peak at large detuning) is compared to the sum of both the peak intensities. This is shown as a function of detuning for various dephasing rates. For zero dephasing the emitter peak becomes dominant as the detuning is increased. It can thus be shown that for zero dephasing and in the limit of large detuning the relative cavity peak intensity scales as $\gamma^2/(\gamma^2+\kappa^2)$, which is very small for typical parameters. In contrast, when dephasing is included, the cavity emission peak is seen to become significant and eventually dominant. Close to resonance the inclusion of dephasing merges the peaks into a single peak. The relative peak intensity is not defined for a single peak and thus not included in fig. \ref{fig:Emission} for $\gamma_p = 10$~GHz and small detuning values.
The cavity emission intensity compared to the total emission intensity is important for the efficiency of the device and we illustrate this in fig. \ref{fig:Emission} (b), where the ratio $\int d\Omega S_C / \int d\Omega \bp{S_C + S_E}$ is shown for varying detuning and dephasing. When $\gamma_p$ is zero and the system is strongly coupled, i.e. when $g_0^2 > \bp{(\kappa-\gamma)/2}^2$ \cite{Andreani99,Rudin99}, most of the light is emitted from the cavity, but for increasing detuning the coupling is weakened and the emission directly from the emitter becomes increasingly important. On resonance an increase in dephasing rate leads to a monotonous decrease in $\int d\Omega S_C$ compared to the total output, and for high dephasing the majority of light is emitted from the emitter.

At zero dephasing and when the detuning is increased the emitter emission becomes more significant. For fixed $\left| \Delta \right| > 0$ an intermediate region appears, where the relative cavity emission displays an increase with dephasing before decreasing toward zero.

\begin{figure}
	[tb]
	\begin{center}
		\includegraphics[width=4.7cm]{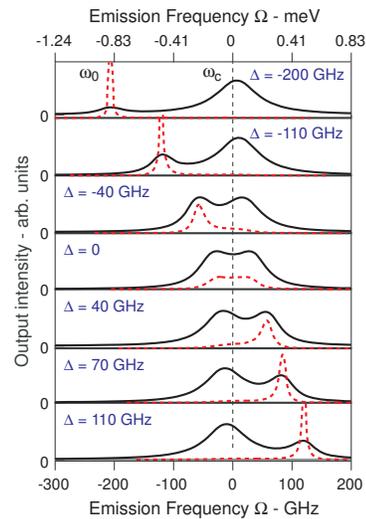} 
        \caption{\label{fig:SpectrumCavityReithmaier}(Color online) Emission spectra calculated using parameters from Reithmaier et al. \cite{Reithmaier04}. The zero dephasing spectra (red dashed line) are downscaled 5 times compared to the $\gamma_p$ = 20 GHz spectra (black line). \textit{Parameters:} $g_0=38$~GHz, $\kappa=43$~GHz, $\gamma=0.1$~GHz.}
	\end{center}
\end{figure}

For high dephasing rates $S_C$ consists of a single peak at the cavity frequency, but the relative cavity emission intensity is smaller compared to zero dephasing. This has to be kept in mind when comparing to measurements, since the distinction between cavity and emitter emission may depend on the experimental set-up and the cavity structure. Before discussing the underlying physics of the intensity shifting effect, let us compare the results of our model to recently published measurements showing a, so far, unexplained detuning dependence.
As an example, fig. \ref{fig:SpectrumCavityReithmaier} shows emission spectra from the cavity, $S_C$, calculated using parameters comparable to the experiments by Reithmaier et al. \cite{Reithmaier04} for different detunings. The measured light is expected to be dominated by $S_C$ because of the high directionality of the micropillar setup.
The spectra including dephasing show much better agreement with the experiment than the spectra calculated in the absence of dephasing. In particular, we note that dephasing favors emission at the cavity frequency although the QD resonance may be far detuned from the cavity resonance. In the experiment \cite{Reithmaier04} the detuning is varied by changing the temperature. It is well known that the dephasing rate is dependent on temperature \cite{Bayer02}, but we emphasize that the enhancement of the cavity peak is robust with respect to variations in the dephasing rate, which is why a fixed $\gamma_p = 20$~GHz is chosen for all values of detuning. This is also the case for different combinations of the decay rates $\kappa$ and $\gamma$ and the intensity shifting effect is present both in the weak and strong coupling regime as noted above. The model has also been tested against data from Yoshie et al. \cite{Yoshie04} and Hennessy et al. \cite{Hennessy07}, and in both cases the unexpected, large emission at the cavity frequency can be explained as an effect of intensity shifting. We notice, however, that the model may be less applicable for these structures.

The simplicity of the model makes the results applicable to a range of systems beyond single photon sources, where two-level systems are coupled to microcavities. An example of this is the work by Strauf et al. \cite{Strauf06} where a few quantum dots were coupled to a nanocavity to realize a photonic-crystal laser. Lasing was witnessed even with the QDs being off resonance with the cavity mode which is surprising and suggests the influence of an effect such as intensity shifting.

\begin{figure}
	[!bt]
	\begin{center}
        \includegraphics[width=7cm]{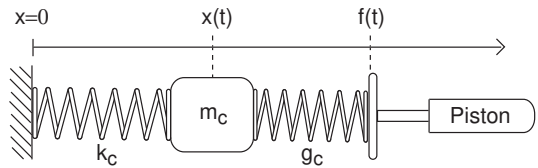}
        \caption{\label{fig:MechSys}Schematic of the mechanic model system. The mass $m_c$ is connected to the wall at $x=0$ through a spring with force constant $k_c$ and to the piston through a spring with force constant $g_c$. The position of the piston is given as $f\bp{t}$.}
   	\end{center}
\end{figure}

In order to get a better physical understanding of the effects responsible for the intensity shifting, we draw upon a mechanical analogue to the QD-cavity system. The differential eqns. (\ref{IV_diff_E1}) and (\ref{IV_diff_C1}) are equivalent to the equations describing a system of two masses, each connected by springs to a wall and mutually coupled by another spring. The resonance frequencies of the uncoupled systems are governed by the masses and the spring constants. For identical spring constants the high detuning limit corresponds to one of the masses being much larger than the other and this mass can then be replaced by a driven piston, which makes the system simpler to analyze and understand. This model is illustrated in fig. \ref{fig:MechSys}. Dephasing events can be thought of as (instantaneously) moving the piston to a new position while keeping the position of the mass fixed (as well as the total energy of the system). In the case of high detuning the equations reduce to
\begin{eqnarray}
\partial_t^2 x\bp{t} + \kappa_c \partial_t x\bp{t} +\bp{k_c+g_c} x\bp{t} = g_c f\bp{t} \label{eq:MechSys}
\end{eqnarray}
where $k_c$ and $g_c$ are force constants for the springs, $\kappa_c$ is the damping of the oscillation and $x\bp{t}$ and $f\bp{t}$ are the position of the mass and the piston, respectively. The mass $m_c$ has been set to unity.
The general solution is the sum of the homogeneous and the inhomogeneous solution, where the former is the damped oscillation of the isolated mass, while the latter is an oscillation at the frequency of the piston. Therefore, the general solution starts out as a combination of the homogeneous and the inhomogeneous oscillation, but over time the transient homogeneous oscillation diminishes and the system oscillates at the frequency of the piston.

Whenever a dephasing event changes the position of the piston, the oscillation of the mass acquires a homogeneous component to compensate for the change. Therefore the mass will acquire a stronger component at its eigenfrequency as the dephasing rate increases, corresponding to a shift in the intensity of the peaks in the Fourier spectrum of the oscillation.

The analogy with the mechanical model demonstrates that the intensity shifting effect is a property of classical as well as quantum mechanical coupled oscillators and the mechanical description of the intensity shifting effect also applies to the quantum mechanical system. At a given time the QD-cavity system is in a superposition of the cavity and emitter state, but the evolution can be changed by a dephasing event, in which case the system must first undergo transient oscillations at the cavity frequency before steady-state oscillation is reestablished.

We note that we have employed the usual assumption that the bare emitter state is excited by a carrier at $t=0$, i.e. $E(0)=1$ \cite{Carmichael89, Cui06}. However, in a more detailed approach one should calculate the excitation of the coupled emitter-cavity states based on the physical excitation of carriers in the system, e.g. off-resonant or near resonant.

In summary, we have investigated a coupled system of a two-level emitter and a cavity and found that the frequency of the emitted light shows a surprising dependence of the dephasing rate. Dephasing shifts the emission intensity towards the cavity frequency, which can explain recent experimental results \cite{Reithmaier04, Yoshie04, Hennessy07}. The intensity shifting effect can  be qualitatively explained by considering the cumulative effect of many dephasing events at a high dephasing rate. The discontinuous phase change adds transients at the cavity frequency to the oscillation, not unlike the ringing effects seen in classical oscillations, and this gives components at the cavity frequency to the emission spectrum. Other effects may of course contribute to the measured spectra. For example the emitter may not be truly two-level, e.g. due to many-body effects, and a more detailed account of the coupling to leaky electromagnetic modes and their emission pattern may need to be given. In general we believe that the results presented are relavant for a wide range of systems and that the intensity shifting effect due to dephasing is of a generic nature and of general relevance to semiconductor systems, which generally are characterized by high rates of dephasing. 

The authors acknowledge helpful discussions with P.~Lodahl and P.~Kaer~Nielsen, Dept. of Photonics Engineering, Technical University of Denmark, and Christian Flindt, Lukin Lab, Department of Physics, Harvard University.

\end{document}